\documentclass[twocolumn,twoside,slac_two]{revtex4}
\usepackage{graphicx}
\usepackage{fancyhdr}
\begin{document}

\title{Fermi-LAT study of two gamma-ray binaries, HESS J0632+057 and AGL J2241+4454}

%

\author{Masaki Mori}
\affiliation{Department of Physical Sciences, Ritsumeikan University, Kusatsu, Shiga 525-8677, Japan}
\author{Akiko Kawachi}
\affiliation{Department of Physics, Tokai University, Hiratsuka, Kanawagwa 259-1252, Japan}
\author{Shigehiro Nagataki}
\affiliation{Yukawa Institute for Fundamental Physics, Kyoto University, Kyoto, Kyoto 606-8502, Japan}
\author{Tsuguya Naito}
\affiliation{Faculty of Management Information, Yamanashi Gakuin University, Kofu, Yamanashi 400-8575, Japan}

\begin{abstract}
GeV gamma-ray emission from two gamma-ray binary candidates, HESS J0632+057 
and AGL J2241+4454, which were recently reported by H.E.S.S. and AGILE, 
respectively, have been searched for using the Fermi-LAT archival dataset. 
Spatial and temporal distribution of gamma-ray events are studied,
but there was no evidence for GeV gamma-ray signal from either sources. 
\end{abstract}

\maketitle

\thispagestyle{fancy}


\section{INTRODUCTION}
X-ray binaries are rather common Galactic X-ray objects and about 
300 sources are catalogued. Recently several objects have been 
reported to emit gamma-rays of GeV and/or TeV energies which are 
modulated in their orbital periods, and a new category of gamma-ray 
binaries is emerging (\cite{Mir2012}), but their emission mechanism is not 
understood well. 
It is clear that we need more observations and samples for the 
detailed study of their nature.

In this study, gamma-ray emissions from HESS J0632+057, for which 
321-day period has been found recently, and AGL J2241+4454, which 
could be identified with a Be star binary with 60-day period, have 
been searched for using the  Fermi-LAT data in the GeV energy range.

\section{HESS J0632+057}

\begin{figure}
\includegraphics[width=80mm]{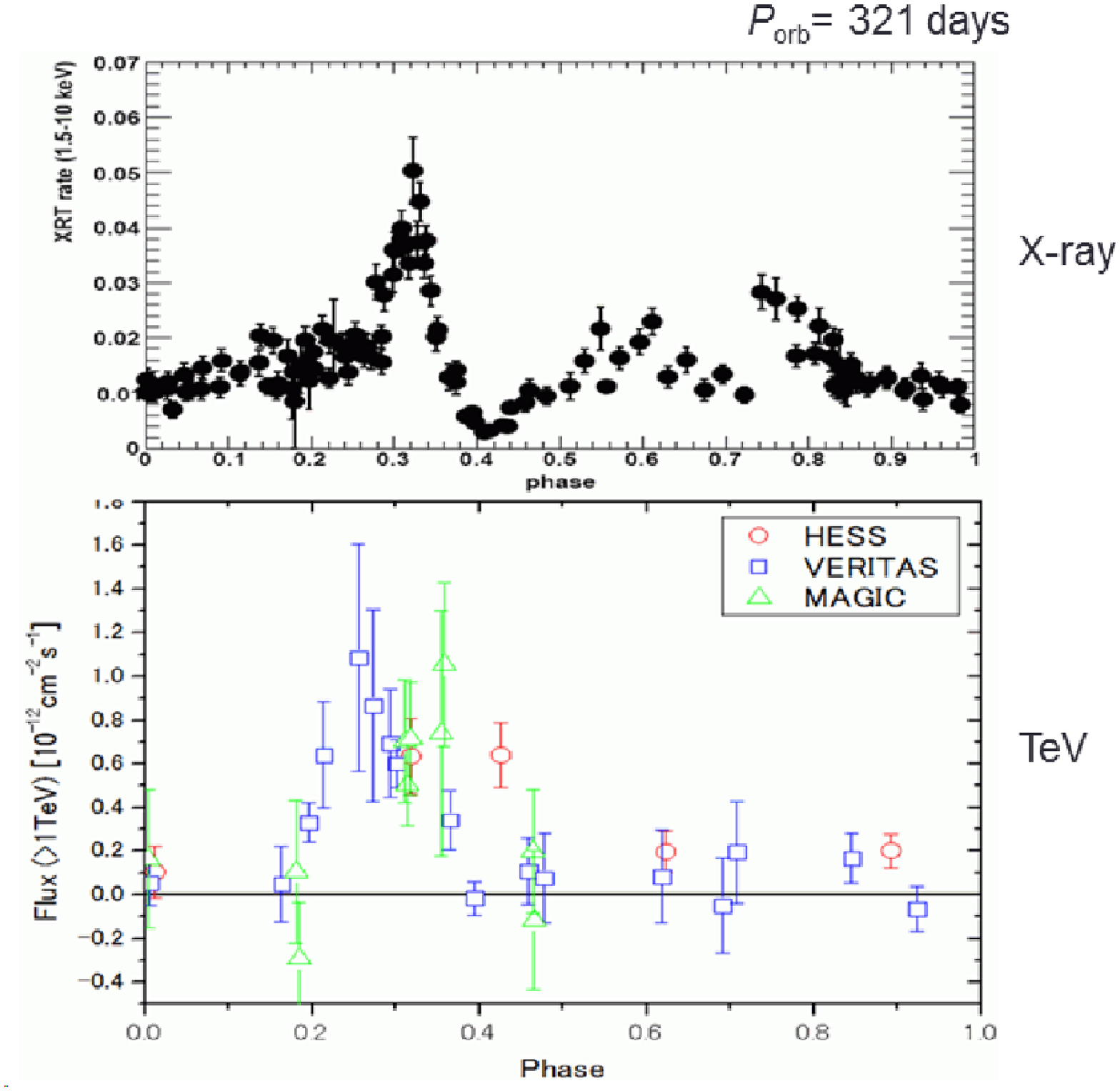}%
\caption{Light curves of HESS J0632+057 (\cite{Mai2011,Ale2012}). \label{fig:J0632lc}}
\end{figure}

This object was found as a TeV point source by H.E.S.S. 
in the Monoceros SNR/Rosetta Nebula region (\cite{Aha2009}). 
It coincides with a massive star MWC148/XMMU J063259.3+054801  
($d\sim 1.5$~kpc) which is variable on hour timescales, and is suspected 
to be a binary system (\cite{Hin2009}). 
Then, $321\pm 5$ day period was found in the XMMU source (\cite{Bon2011}). 
TeV follow-up observations for 6 years have revealed that 
gamma-ray fluxes  are modulated at this period (\cite{Mai2011,Ale2012}) 
(see Fig.\ref{fig:J0632lc}). 
However, this object is not listed in the Second Fermi-LAT catalog
(\cite{Nol2012}).

\begin{figure}
\includegraphics[width=80mm]{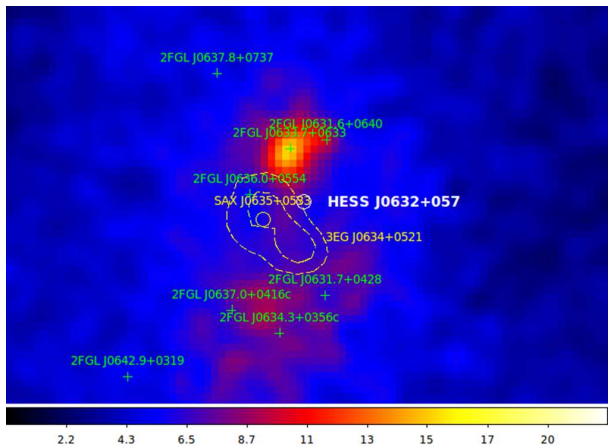}%
\caption{Gamma-ray countmap above 200 MeV around HESS J0632+057. \label{fig:J0632cm}}
\end{figure}

\begin{figure}
\includegraphics[width=80mm]{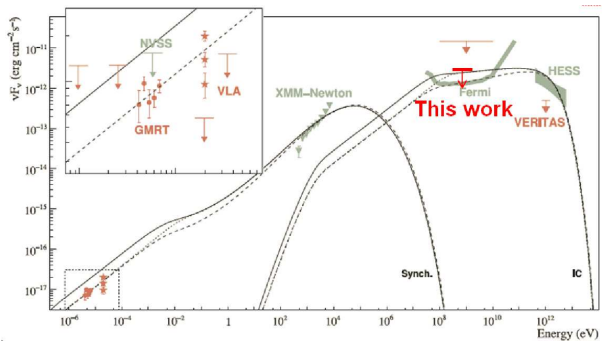}%
\caption{Multiwavelength spectrum of HESS J0632+057 (\cite{Hin2009})
with our upper limit. See text for model curves, \label{fig:J0632sp}}
\end{figure}

We analyzed LAT gamma-ray data above 200 MeV for 3.5 years 
using the Fermi Science Tools (version v9r27p1) with {\tt P7SOURCE\_V6}
response function. 
The resulting skymap is shown in Fig. \ref{fig:J0632cm}. 
The likelihood analysis yielded no significant signal, and
we found no evidence for gamma-ray emission. We obtained 
an upper limit of $1.0\times 10^{-8}$cm$^{-2}$s$^{-1}$ (90\% C.L.) 
for 1-year data (3.5 year-data is under analysis). 
The spectral energy distribution is shown in Fig. \ref{fig:J0632sp} 
with radio and X-ray data (\cite{Ski2009}) 
where the GeV limit is very close to model expectations 
assuming inverse Compton emission (solid: $E^{-2.0}$ electron injection 
with $E_{\rm min}= 1$~GeV, dashed: $E^{-1.9}$ / 1 GeV, dotted: $E^{-2}$ / 2 GeV).

\begin{figure}
\includegraphics[width=80mm]{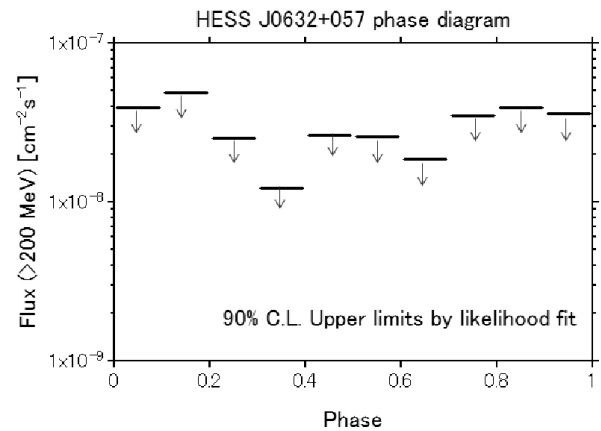}%
\caption{Orbital-phase-resolved upper limits on HESS J0632+057. \label{fig:J0632op}}
\end{figure}

Orbital modulation of GeV gamma-ray emission has been investigated 
assuming the 321 day period. We could not find any significant phase bins, 
and Fig. \ref{fig:J0632op} shows the upper-limit light curve.

\section{AGL J2241+4454}

AGILE reported the discovery of this object for a short period 
(2010-07-25/26) with a flux of $1.5\times 10^{-6}$cm$^{-2}$s$^{-1}$ 
above 100 MeV (\cite{Luc2010}). 
However, Fermi-LAT observations could not confirm this detection 
and set an upper limit of $1.0\times 10^{-7}$cm$^{-2}$s$^{-1}$ (95\% C.L.) 
above 100 MeV (\cite{Fer2012}). 
It could be identified as a Be star HD 215227 (MWC 656) showing 
an orbital period of $60.37\pm 0.04$ days (\cite{Wil2010}).

We analyzed LAT gamma-ray data above 100 MeV using the Fermi Science Tools
as in the previous section.
The resulting skymap is shown in Fig.5 (2010-07-25/26) 
and Fig. 6 (3.5 years). 
We found no evidence for gamma-ray emission and obtained 
upper limits of $7.2\times 10^{-8}$cm$^{-2}$s$^{-1}$ and 
$9.4\times 10^{-10}$cm$^{-2}$s$^{-1}$ (90\% C.L.) 
for the two-day data and 3.5-year data,  respectively.

\begin{figure}
\includegraphics[width=80mm]{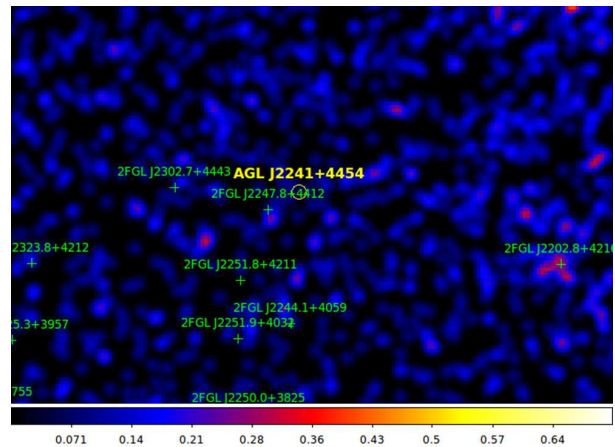}%
\caption{Gamma-ray countmap above 100 MeV around AGL J2241+4454 (2010-07-25/26). 
\label{fig:j2241cm}}
\end{figure}

\begin{figure}
\includegraphics[width=80mm]{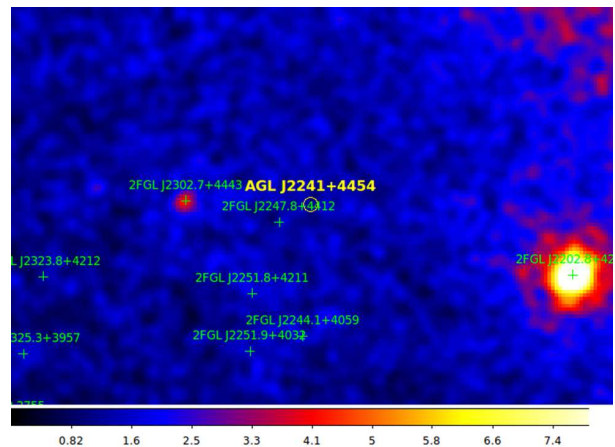}%
\caption{Gamma-ray countmap above 100 MeV around AGL J2241+4454 (3.5 years). 
\label{fig:j2241cm2}}
\end{figure}

\begin{figure}
\includegraphics[width=80mm]{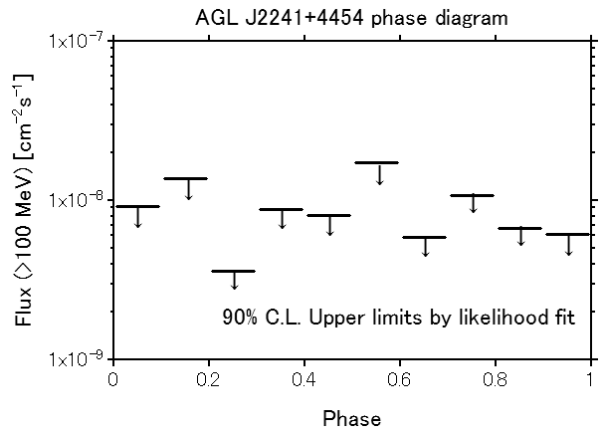}%
\caption{Orbital-phase-resolved upper limits on AGL J2241+4454. \label{fig:j2241op}}
\end{figure}

Orbital modulation of GeV gamma-ray emission has been investigated 
assuming the 60.37 day period. We could not find any significant 
phase bins, and Fig. \ref{fig:j2241op} shows the upper-limit light curve.

\section{SUMMARY}

We have searched for GeV gamma-ray emission from HESS J0632+057 
and AGL J2241+4454 using the Fermi-LAT data. 
No significant signal was found from either objects and long-term 
and orbital-phase-resolved upper limits have been set on gamma-ray 
fluxes which set restriction on their high-energy activities.

\bigskip 
\begin{acknowledgments}
This work was supported by JSPS KAKENHI Grant Number 22540315 (MM).
\end{acknowledgments}

\bigskip 

\end{document}